\documentclass[12pt,preprint]{aastex}

\slugcomment{\today}

\def\ea{{\it et al.\,}}
\def\be{\begin{equation}}
\def\ee{\end{equation}}

\begin{document}
\title{COSMIC MICROWAVE BACKGROUND TEMPERATURE AT GALAXY CLUSTERS}

\author{E.S. Battistelli, M. DePetris, L. Lamagna, F. Melchiorri, E. Palladino, G. Savini}
\affil{ Department of Physics, University "La Sapienza", P.le A.
Moro 2, 00185, Rome, Italy}

\author{A. Cooray}
\affil{Theoretical Astrophysics, California Institute of
Technology, Pasadena CA 91104}

\author{A. Melchiorri}
\affil{University of Oxford, Denys Wilkinson building,
Astrophysics, Keble Road, Oxford, OX1-3RH, UK}

\author{Y. Rephaeli, M. Shimon}
\affil{School of Physics and Astronomy, Tel Aviv University}

\begin{abstract}

We have deduced the cosmic microwave background (CMB) temperature
in the Coma cluster (A1656, $z=0.0231$), and in A2163 ($z=0.203$)
from spectral measurements of the Sunyaev-Zel'dovich (SZ) effect
over four passbands at radio and microwave frequencies. The resulting
temperatures at these redshifts are
$T_{Coma} = 2.789^{+0.080}_{-0.065}$ K and $T_{A2163} =
3.377^{+0.101}_{-0.102}$ K, respectively. These values
confirm the expected relation $T(z)=T_{0}(1+z)$, where $T_{0}=
2.725 \pm 0.002$ K is the value measured by the COBE/FIRAS experiment.
Alternative scaling relations that are conjectured in non-standard
cosmologies can be constrained by the data; for example, if $T(z)
= T_{0}(1+z)^{1-a}$ or $T(z)=T_{0}[1+(1+d)z]$, then
$a=-0.16^{+0.34}_{-0.32}$ and $d = 0.17 \pm 0.36$ (at 95\%
confidence). We briefly discuss future prospects for more precise
SZ measurements of $T(z)$ at higher redshifts.
\end{abstract}

\keywords{cosmic microwave background---cosmology:observations---
galaxies:clusters:individual(A1656,A2163)}

\section{Introduction}

The (present) CMB temperature was precisely measured by the Far
Infrared Absolute Spectrometer (FIRAS) on board the COBE
satellite, $T_{0} = 2.725 \pm 0.002 \ $K, in the frequency range
2-20 cm$^{-1}$ \cite{mather1999}. These measurements essentially
rule out all cosmological models in which the CMB spectrum is
non-Planckian at $z=0$. Models with a purely blackbody spectrum
but with a different $T(z)$ dependence than in the standard model
are, however, unconstrained by the FIRAS database. Also
unconstrained are models with spectral distortions that are now
negligible, but were appreciable in the past. A specific example
is the relation $T(z) = T_{0}(1+z)^{1-a}$, where $a$ is a
parameter of the theory (see, e.g., Lima et al. 2000). More
generally, models in which ratios of some of the fundamental
constants vary over cosmological time are also of considerable
interest.

So far, $T(z)$ has been determined mainly from measurements of
microwave transitions in interstellar clouds which contain atoms
and molecules that are excited by the CMB (as reviewed by
LoSecco \ea 2001). The temperature has been determined in the
Galaxy, as well as in clouds at redshifts up to $z \sim 3$
(Levshakov \ea 2001). These measurements are affected by
substantial systematic uncertainties stemming from the unknown
physical conditions in the absorbing clouds (Combes and Wiklind
1999).

The possibility of determining $T(z)$ from measurements of the
Sunyaev-Zel'dovich (SZ) effect had been suggested long ago
(Fabbri, Melchiorri \& Natale 1978, Rephaeli 1980). (For general
reviews of the effect and its cosmological significance, see
Rephaeli 1995a, Birkinshaw 1999.) The proposed method is based on
the steep frequency dependence of the change in the CMB spectral
intensity, $\Delta I$, due to the effect, and the weak dependence of
ratios $\Delta I(\nu_{i}) / \Delta I(\nu_{j})$ of intensity changes
measured at two frequencies ($\nu_{i},\, \nu_{j}$) on properties of
the cluster (Rephaeli 1980). Because of this, and the fact that -- in
the standard cosmological model -- the effect is essentially
independent of $z$, SZ measurements have the potential of yielding
much more precise values of $T(z)$ than can be obtained from ratios
of atomic and molecular lines. With the improved capability of
reasonably precise spectral measurements of the SZ effect, the method
can now be used to measure $T(z)$ in nearby and moderately distant
clusters. Here we report first results from spectral analysis of SZ
measurements in the Coma and A2163 clusters of galaxies.

\section{$T(z)$ from SZ}

The CMB intensity change due to Compton scattering in a cluster
can be written in the form
\be
\Delta I = { 2 k^{3} T^{3} \over h^{2}c^{2} } {x^4e^x \over (e^x
-1)^2} \int d\tau \biggr[\theta f_{1}(x) - \beta + R(x, \theta,
\beta) \biggl ] \, , \ee where $x = h\nu /kT$ is the
non-dimensional frequency, $\theta = kT_e /mc^2$ is the electron
temperature in units of the electron rest energy, and $\beta$ is
the line of sight (los) component of the cluster (peculiar)
velocity in the CMB frame in units of $c$. The integral is over
the Compton optical depth, $\tau$. Both the thermal (Sunyaev \&
Zeldovich 1972) and kinematic \cite{SunZel80} components of the
effect are included in equation (1), separately in the first two
(additive) terms, and jointly in the function $R(x,\theta,\beta)$.
In the non-relativistic limit (which is valid only at low electron
temperatures and frequencies) the spectral dependence of $\Delta
I$ is fully contained in the product of the $x$-dependent
pre-factor times the function $f_{1}(x) = x(e^x +1)/(e^x -1) - 4$.
The more exact treatment of Compton scattering in clusters
necessitates a relativistic calculation (Rephaeli 1995b) due to
the high electron velocities. The function $R(x,\theta,\beta)$
includes the additional spectral, temperature, and (cluster)
velocity dependence that is obtained in a relativistic treatment.
This function can be approximated by an analytic expression that
includes terms to orders $\theta ^5$ and $\beta^{2} \theta$:
\be
R(x, \theta, \beta) \simeq \theta^{2} \biggr[f_{2}(x)+\theta
f_{3}(x) + \theta^{2}f_{4}(x) + \theta^{3}f_{5}(x) \biggl] - \beta
\theta \biggr[ g_{1}(x) + \theta g_{2}(x) \biggl] +
\beta^{2}\biggr[1 + \theta g_{3}(x) \biggl] \,. \ee The spectral
functions $f_{i}$  and $g_{i}$ were determined by Itoh \ea (1998),
Itoh \ea (2002), and Shimon \& Rephaeli (2002). For our purposes
here this analytic approximation is sufficiently exact even close
to the crossover frequency. The non-relativistic limit, $R \equiv
0$, applies if the sum of all these terms can be ignored at the
desired level of accuracy.

The $z$ dependence of $\Delta I$ is fully determined by the
functions $\nu = \nu (z)$, and $T = T(z)$.
The temperature-redshift relation may assume various forms in
non-standard cosmologies; here we consider two examples. In the
first, $T(z) = T(0)(1+z)^{(1- a)}$, where $a$ is taken to be a free
parameter, but with the standard scaling $\nu = \nu_{0}(1+z)$
unchanged. With these relations the non-dimensional frequency
obviously depends on $z$, $x = x_{0}(1+z)^{a}$, if $a \neq 0$;
here, $x_{0}= h\nu_{0}/kT(0)$. Another functional form which seems
also to be of some theoretical interest is $T(z) = T(0)[1+(1+d)z]$
(LoSecco \ea 2001), for which $x =x_{0}(1+z)/[1+(1+d)z]$. Obviously,
in the standard model $a=d=0$.

For a slow moving ($\beta < 10^{-3}$) cluster, the expression
for $\Delta I$ in the non-relativistic limit depends linearly
on the Comptonization parameter, $y = \int \theta d\tau$, which
includes all dependence on the cluster properties. A ratio of
values of $\Delta I$ at two frequencies is then essentially
independent of these cluster properties. In the more general
case, the first term in the square parentheses in eq. (1) still
dominates over the other two, except near the crossover frequency
(whose value generally depends on $T_e$, except in the
non-relativistic limit where $x_c = 3.83$; Rephaeli 1995b,
Nozawa \ea 1998, Shimon \& Rephaeli 2002), where the sum of the
temperature dependent terms vanish. For values of $x$ outside
some range (roughly, $3.5 < x < 4.5$), the dependence of
$\Delta I$, and -- particularly, a ratio of values of
$\Delta I$ -- on $\beta$ is very weak since
the observed temperature range in clusters corresponds to
$0.006 < \theta < 0.03$, whereas typically $\beta < 0.002$.

\section{Data Analysis}

We have analyzed results of SZ measurements in the Coma cluster
(A1656) and A2163. Measurements of Coma, $z= 0.0231 \pm 0.0017$
\cite{struble1999} were made with the MITO \cite{depetris2002}
telescope in 20 hours of integration. We also use the result of
measurements at 32 GHz made with the OVRO 5.5m telescope (Herbig
\ea 1995, Mason \ea 2001). A2163, $z=0.203\pm 0.002$ \cite{arnaud1992},
was observed with the SuZIE array (Holzapfel \ea 1997a), and with
both the OVRO and BIMA interferometric arrays (LaRoque \ea 2000).

When observing a cluster the SZ part of the measured signal is
\be
\Delta S_i= G_i A\Omega|_i \int_0^\infty \Delta I(0) \epsilon_i
(\nu) d\nu \, , \ee where $G_i$ is the responsivity of the
$i^{th}$ photometric channel, $A\Omega|_i$ is the corresponding
throughput, and $\epsilon_i(\nu)$ is the spectral efficiency. The
full measured signal includes also contributions from the
atmosphere, CMB anisotropies, and -- at very high frequencies --
also emission from dust. Multifrequency observations allow us to
remove contributions from both the primary CMB anisotropy and the
kinematic SZ effect, as has been attempted in the analysis of MITO
measurements of the Coma cluster (DePetris \ea 2002).

The ratio of signals in two different photometric channels $i$ and
$j$ is
\be
{\Delta S_i\over \Delta S_j} = {G_i\over G_j} {A\Omega|_i\over
A\Omega|_j} {\int_0^\infty {{x}^4 e^{x} \over (e^{x}-1)^2}
\biggr\{\int d\tau \biggr[\theta f_{1}(x) -\beta +
R(x,\theta,\beta) \biggl]\biggl\} \epsilon_i(\nu) d\nu\over
\int_0^\infty{{x}^4 e^{x} \over (e^{x}-1)^2} \biggr\{\int d\tau
\biggr[\theta f_{1}(x) -\beta +
R(x,\theta,\beta)\biggl]\biggl\}\epsilon_j(\nu) d\nu} \, . \ee The
main dependence on the cluster properties in $y$ cancels out when
$\beta$ is negligible. Multi-frequency observations that include
measurements at the crossover frequency (e.g., MITO) afford
effective separation of the thermal and kinematic components,
exploiting their very different spectral shapes. This was
demonstrated in the analysis of MITO measurements of the Coma
cluster (De Petris et al. 2002). Since the above ratio depends
weakly on the cluster velocity, the residual uncertainty due to
velocities of even $\pm 500$ km/s can be ignored in comparison
with other errors. The ratio is also weakly dependent on the gas
temperature at a level which we found to correspond to $\sim 1$\%
uncertainty in the estimation of the CMB temperature (for a
typical observational error in $T_e$). Moreover, the uncertainty
associated with the absolute calibration, $G_i$, is largely
removed once we fit data from several photometric channels, as
long as they are calibrated with a source with a known spectrum
(e.g., a planet) even if its absolute calibration is uncertain.
Only relative uncertainties among the various spectral channels
are important; these include differences in angular response and
in atmospheric transmittance. A standard blackbody source with a
precisely calibrated temperature is therefore not required. We
expect these considerations to imply that the precision of CMB
temperature measurements via the SZ effect will not be appreciably
affected by most of the known systematic errors. The level of
precision in the measurement of $T(z)$ is limited largely by other
observational uncertainties, as discussed below.

The responsivity $G_i$ of each channel is usually determined from
detailed observations of very well measured sources such as
planets (mostly Jupiter or Saturn) and the Moon. While the
temperature uncertainty of these sources can be as large as
10$\%$, their spectra are relatively well known. For a source at a
temperature of $T_S$ and with a throughput $\Omega_S$, the signal
in the Rayleigh-Jeans part of the spectrum is
\be
\Delta S_i= G_i A_i\Omega_S T_S{2k\over c^2}\int_0^\infty
\nu^2\epsilon_i(\nu) \eta(\nu) d\nu \, .
\ee
Since we are interested in the ratio of the signals in two
channels, $T_S$ drops out from the final expression, and the
uncertainty in the ratio of values of the brightness temperature
in the two channels depends only on the emissivity of the planet,
$\eta(\nu)$, convolved with the spectral efficiencies of the
channels, $\epsilon_i(\nu)$. Note that the throughputs of individual
channels, $A\Omega|_i$, are usually made almost equal by an
appropriate choice of the optical layout.

The main uncertainty in the MITO measurements is due to imprecise
knowledge of the bandwidth and transmittance of the filters.
Therefore, we have precisely measured the total efficiency of our
photometer by means of a lamellar grating interferometer. The
response of the photometer is measured when it is illuminated with
a laboratory blackbody at different temperatures ranging from
liquid $N_2$ to room temperature. The effect of the uncertainty in
our bandwidths has two main consequences. The first is the error
in the ratio of responsivities of channel $i$ and channel $j$,
$G_i/G_j$. We estimate this error by computing the ratio of
expected signals evaluated by convolving Moon and Jupiter spectra
with the spectra of our filters, and taking into account
spectroscopic uncertainties; we obtain a final error estimate
which is less than 1$\%$. The second consequence of bandwidth
uncertainty enters in the integrations over the SZ spectral
functions in eq.[4]. However, this latter error is smaller than
the former and is practically negligible.

Also quite precisely measured were the angular responses of the
telescope and the photometer in the four channels. We have studied
differences in the response of the four channels to both a point
source, such as Jupiter, and an extended source, such as the Moon,
as they cross the field of view. The optical layout of our
photometer is such that all four detectors observe the same sky
region. The differences among channels are less than 1\% for
extended sources. Finally, we have also studied the change in
relative efficiency of our observations under different
atmospheric conditions. The simulations show a change of $\sim 1$
mK in the estimated CMB temperature if the water vapor content
changes from 0.5 mm to 1.5 mm, as determined by convolving
theoretical atmospheric spectra with our filters. The
uncertainties in determining the water vapor content, related to
our procedure of measuring the atmospheric transmittance through a
sky-dip technique, are at a level of 30$\%$. However, when the
effect of this error is propagated through our procedure and
analysis we estimate that it amounts to a negligible error in the
final result for the CMB temperature.

In conclusion, systematic errors contribute less than 3\% to the
observed MITO signals, and are quite negligible with respect to
the 10$\%$ contribution associated with the residual noise of the
four detectors. For the OVRO measurement, we use the total error
as specified by Mason \ea (2001); the statistical weight of this
low frequency measurement is such that it only contributes $\simeq
12$\% to the final results.

Our assessment of the errors in the SuZIE and OVRO \& BIMA
measurements of A2163 (LaRoque \ea 2000) is far less certain than
those in the MITO measurements, since we do not know the exact
spectral responses of their filters and the atmospheric conditions
during each of these observations. We have used Gaussian profiles
for the SuZIE filters, with peak frequencies and FWHM values as
given by Holzapfel \ea (1997a). For the OVRO \& BIMA data point,
we also take a Gaussian profile with a peak frequency at 30 GHz
and a 1 GHz bandwidth. We have employed data provided by SuZIE,
for which a correction due to thermal dust emission has been taken
into account. The Coma and A2163 data are collected in Table 1 and
Table 2. Finally, the value we used for the electron temperature
is $(8.25\pm0.10)$ keV, as measured with XMM in the central region
of Coma \cite{arnaud2001}; the value for A2163 is
$12.4^{+2.8}_{-1.9}$ keV \cite{holza97b}.

In both sets of cluster observations, we have minimized the
difference between theoretical and experimental ratios (properly
weighted by statistical errors) with $T$ as a free parameter. The
results are $T_{Coma}(0) = 2.726^{+0.078}_{-0.064}$ K which,
when multiplied by $(1+z)$ with $z= 0.0231 \pm 0.0017$, yields
\be
T_{Coma} = 2.789^{+0.080}_{-0.065} \rm{K} \, .
\ee
The main uncertainty is in the SZ observations, with only a small
error ($7\times 10^{-4}$ K) due to the uncertainty ($0.1$ keV) in the
electron temperature. For A2163 we obtain $T_{A2163}(0) =
2.807^{+0.084}_{-0.085} \ K $, and when multiplied by $(1+z)$ with
$z=0.203 \pm 0.002$, yields
\be
T_{A2163} = 3.377^{+0.101}_{-0.102} \ \rm{K} \, .
\ee
The larger error in $T_e$ ($\sim 2$ keV) translates to an uncertainty
of $1.5\times 10^{-2}$ K in the deduced CMB temperature at the
redshift of A2163. The mean of the above two values for $T(0)$ is
$\langle T \rangle= 2.766^{+0.058}_{-0.053}$ K, in good
agreement with the COBE value.

Using the COBE value of the temperature and the two values deduced
here for $T(z)$ at the redshifts of Coma and A2163, we fit the
three data points by the relation $T(z)= T(0)(1+z)^{1-a}$; the
best fit is shown by the dashed line in Figure 1. It corresponds
to $a=-0.16^{+0.34}_{-0.32}$ at 95\% confidence level (CL). With the
alternative scaling $T(z)= T(0)[1+ (1+d)z]$, we obtain
a best fit value of $d=0.17\pm 0.36$ at the 95\% CL.

These first SZ results for $T(z)$ are clearly consistent with
the standard relation, although the most probable values of both
$a$ and $d$ indicate a slightly stronger $z$ dependence. Our
values for these two parameters are quite similar to those deduced
by LoSecco \& Mathews (2001) from measurements of microwave
transitions. Their results -- based on two firm detections at
redshifts $2.338$ and $3.025$ -- are $a=-0.05\pm 0.13$ and $d=
0.10\pm 0.28$ (at 95\% CL). Thus, our SZ-based
values already provide the same level of precision even though the
two clusters are at much lower redshifts ($z \lesssim 0.2$). SZ
measurements of higher redshift clusters will clearly provide a
preferred alternative to the atomic and molecular lines method.
If we fit the above relations to the combined data sets (consisting
of all five temperatures), we obtain $a=-0.09 \pm 0.20$ and $d=0.14\pm
0.28$ (at $95\%$ CL).
Finally, Molaro {\it et al.} (2002) have recently questioned the
validity of the temperature measurement at $z=2.338$ (by LoSecco \&
Mathews 2001), and have revised the value at $z=3.025$ to
$12.1^{+1.7}_{-3.2}$K. We have thus repeated the fit excluding the
fourth data point and using the latter value for the fifth; the
results from this fit are essentially unchanged: $a=-0.11\pm 0.22$,
and $d=0.16\pm 0.32$.

\section{Discussion}

The method employed here to measure $T(z)$ can potentially yield
very precise values which will tightly constrain alternative
models for the functional scaling of the CMB temperature with
redshift, and thereby provide a strong test of non-standard
cosmological models. In order to improve the results reported
here, multi-frequency measurements of the SZ effect in a
significantly larger number ($\sim 20$) of clusters are needed.
The clusters should include nearby ones in order to better
understand and control systematic errors. Exactly such
observations are planned for the next 2-3 years with the upgraded
MITO project (Lamagna \ea 2002), and the stratospheric BOOST
experiment (Lubin P., private communication) and OLIMPO experiment
(Masi \ea 2003). These experiments will employ sensitive
bolometric arrays at four frequency bands with high spatial
resolution.

As detector noise is reduced, the uncertainty in the electron
temperature may contribute the dominant relative error. For
example, an uncertainty of $\sim 2$ keV -- as in the case of Abell
2163 -- becomes dominant when detector noise is reduced to a level
ten times lower than the values used here. Therefore, it will be
necessary to select clusters for which the gas temperature is
precisely (to a level of $\sim 0.1$ keV) known. With the large
number of clusters that have already been -- or will be --
sensitively observed by the XMM and {\it Chandra} satellites, it
should be possible to optimally select the SZ cluster sample.

With the currently achievable level of precision in intracluster
gas temperature measurements, the present technique makes it
possible to determine $a$ and $d$ with an uncertainty that is not
less than $0.03$ on both parameters, even with ideal
(i.e., extremely high sensitivity) SZ measurements. The projected
sensitivities of the future Planck and Herschel satellites will
enable reduction of the overall error in the
values of these two parameters by more than an order of magnitude.
In addition to limits on alternative CMB temperature evolution
models, such a level of precision will open new possibilities of
testing the time variability of physical constants. For example,
it will be feasible to measure a possible variation of the fine
structure constant, thereby providing an alternative to the current
principal method which is based on spectroscopic measurements of
fine structure lines in quasar absorption spectra.

\clearpage

\begin{figure}
\plotone{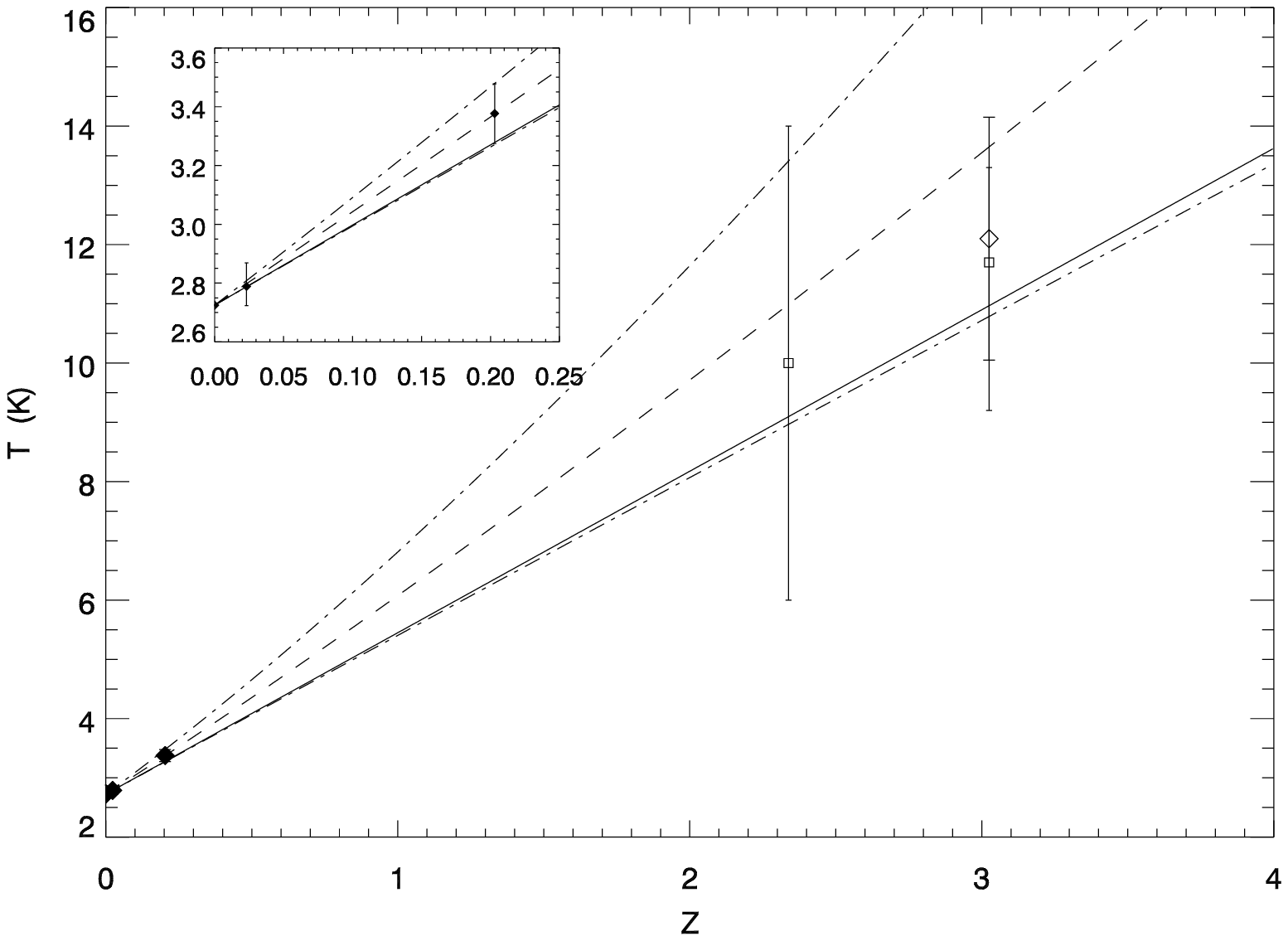} \caption{Black diamonds mark values of the CMB
temperature obtained by COBE/FIRAS and SZ measurements. The two
values reported by LoSecco {\it et al.} (2001) are shown as white
squares, and the white diamond is the value reported by Molaro
{\it et al.} (2002). The solid line is $(1+z)$ scaling law as
predicted in the standard model. The dashed line shows the best
fit to the alternative scaling law with $a$ determined by the
COBE/FIRAS and SZ measurements, whereas dot-dashed lines are the
$\pm 1\sigma$ values of $a$. The inset shows a magnified view of
the region $z \leq 0.25$.\label{fig_1}}
\end{figure}

\clearpage

\begin{deluxetable}{cccrr}
\footnotesize \tablecaption{SZ signals from COMA used in the
present analysis. \label{tbl-1}} \tablewidth{0pt}
\tablehead{\colhead{Experiment} & \colhead{Frequency [GHz]} &
\colhead{Bandwidth [GHz]} & \colhead{$\Delta I\ [MJ sr^{-1}]$} &
\colhead{$\Delta T\ [\mu K]$}} \startdata OVRO &$32.0$&$13.0$
&$-0.0159\pm 0.0028$ &$-520\pm 93$\\ MITO1 &$143$&$30$ &$-0.068\pm
0.014$ &$-179.3\pm 37.8$ \\ MITO2 &$214$&$30$ &$-0.016\pm 0.039$
&$-33.4\pm 81.2$ \\ MITO3 &$272$&$32$ &$0.075\pm 0.015$ &$169.8\pm
35.1$\\
\enddata
\end{deluxetable}

\clearpage

\begin{deluxetable}{cccrr}
\footnotesize \tablecaption{SZ measurements of A2163 used in the
present analysis. \label{tbl-2}} \tablewidth{0pt}
\tablehead{\colhead{Experiment} & \colhead{Frequency [GHz]} &
\colhead{Bandwidth [GHz]}& \colhead{$\Delta I\ [MJ sr^{-1}]$} &
\colhead{$\Delta T\ [\mu K]$}} \startdata OVRO+BIMA &$30$&$1$
&$-0.048\pm 0.006$ &$-1777\pm 222$\\ SuZIE1 &$141.59$&$12.67$
&$-0.380\pm 0.037$ &$-1011.3\pm 98.0$
\\ SuZIE2 &$216.71$&$14.96$ &$-0.103\pm 0.077$ &$-213.0\pm 159.3$
\\ SuZIE3 &$268.54$&$25.66$ &$0.295\pm 0.105$ &$662.2\pm 235.7$\\
\enddata
\end{deluxetable}

\end{document}